\documentclass[ps,prl,10pt,twocolumn,amsmath,amssymb,superscriptaddress,longbibliography,floatfix]{revtex4-2}
\usepackage[urlcolor=blue,colorlinks=true,citecolor=blue,linkcolor=blue,pdfstartview={FitH},bookmarks=false]{hyperref}
\usepackage{graphicx}
\usepackage{xcolor}
\usepackage{hyperref}
\usepackage{adjustbox}
\usepackage{subfigure}

\newcommand{\cdag}{c^{\dagger}}
\newcommand{\cnod}{c^{\phantom{\dagger}}}

\newcommand{\e}{\textrm{e}}

\newcommand{\MD}[1]{\textcolor{black}{#1}}

\begin{document}

\title{Electronic correlations driving Chirality-Induced Spin Selectivity}

\author{Jacek Herbrych}
\affiliation{
Institute of Theoretical Physics,
Faculty of Fundamental Problems of Technology,
Wroc{\l}aw University of Science and Technology,
50-370 Wroc{\l}aw, Poland}

\author{Maria Daghofer}
\email{maria.daghofer@fmq.uni-stuttgart.de}
\affiliation{
Institut f\"ur Funktionelle Materie und Quantentechnologien,
Universit\"at Stuttgart,
70550 Stuttgart, Germany}

\begin{abstract}
 We explicitly account for electron-electron interactions when modeling low-dimensional helical organic molecules. We show that competition between various hopping channels, together with interaction-induced double- and superexchange mechanisms, can stabilize non-collinear helical magnetic order. The resulting single-electron bands exhibit partial spin polarization, a manifestation of $p$-wave magnetism. Using density-matrix renormalization group, cluster perturbation theory, and Monte Carlo methods, we find that even vanishingly small spin-orbit coupling triggers strong spin selectivity at temperatures significantly above the spin-orbit scale. While strong correlations are essential for this mechanism, long-range spin ordering is not required. We thus propose non-collinear spin correlations driven by Coulomb interactions as an explanation of chirality-induced spin selectivity and discuss connections to experiments.
\end{abstract}
\maketitle

\begin{figure}
  \includegraphics[width=1.0\columnwidth]{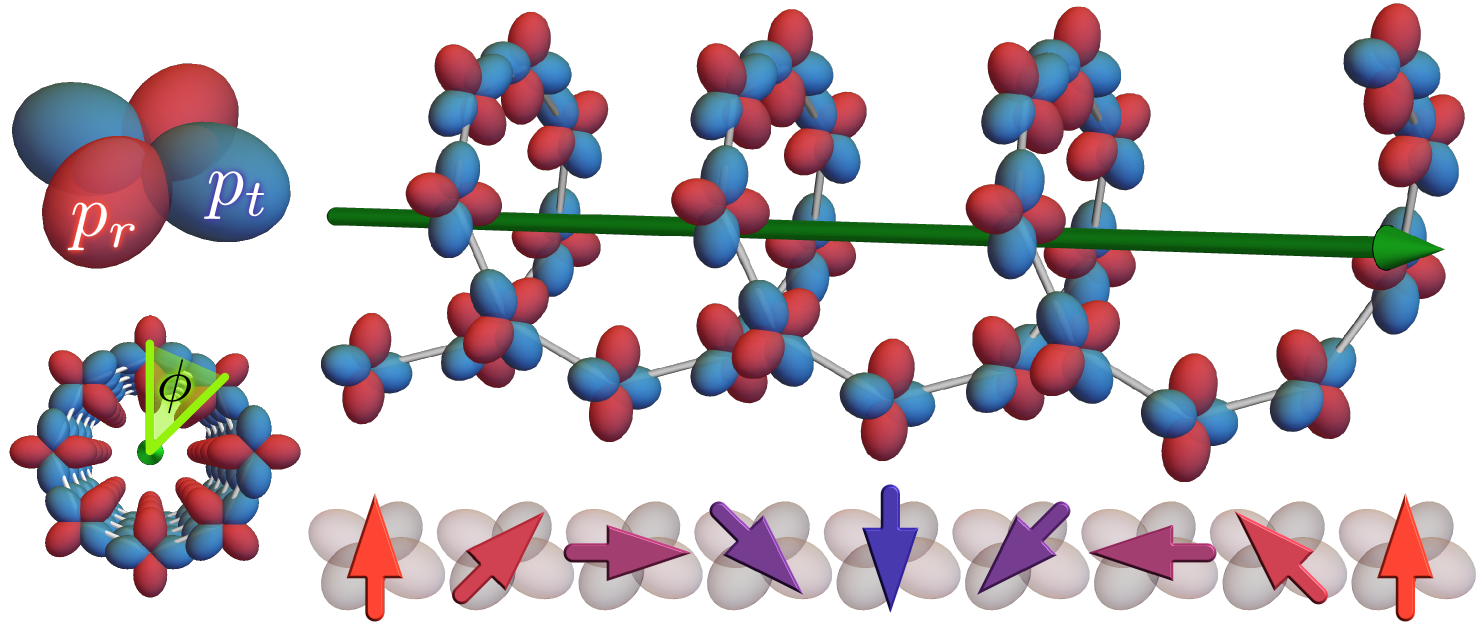}
  \caption{Cartoon of the two-orbital model motivated by Ref.~\cite{chiral_orb_dna}: each site contains 'radial' orbital $p_r$ and 'tangential' $p_t$, see top left. The spiral is shown from the side (top) and from above (bottom left). The spin-spin angle in the correlation-driven spin spiral (bottom right) is set by correlations, while SOC only selects the spin plane (orthogonal to the spiral axis) and the winding direction.}
  \label{fig:cartoon}
\end{figure}

\textit{Introduction ---} 
Chirality-induced spin selectivity (CISS) refers to a transport phenomenon in which electric currents passing through a chiral medium become spin polarized despite the absence of static ferromagnetic (FM) order. The polarization is tied to the chiral structure of the molecules (see sketch in Fig.~\ref{fig:cartoon}), and is particularly evident in DNA, the most iconic helix. Electrons transmitted through DNA can acquire substantial spin polarization, indicating that the molecule itself acts as an efficient spin filter~\cite{Gohler2011}. As spin can affect chemical reactions~\cite{spin_chemistry_15,Rev_NChem_19}, spin-dependent transport has accordingly been proposed to also play a role in biology~\cite{DNA_quantum}. Further, CISS has motivated the exploration of biomolecules as functional elements in molecular spintronics~\cite{Naaman2015}. 

Beyond DNA, CISS appears in a wide variety of systems and often remains remarkably robust even at elevated temperatures~\cite{Review_ACS_24}. It has been observed in biological systems like chiral proteins in cell membranes~\cite{Mishra2020} and peptide-based nanofibers~\cite{nanofiber_20}, as well as in inorganic crystals~\cite{PhysRevLett.124.166602}. A few of its manifestations are: (i) Photo-excited electrons emitted from chiral molecules exhibit strong spin polarization~\cite{Gohler2011,Xie2011}. (ii) Adsorption of chiral molecules can depend on the substrate's magnetization~\cite{deposition_18,Adsorb_Ernst_24}. (iii) Chiral molecules on FM substrates show substantial magnetoresistance~\cite{nanofiber_20,peptide_dna_mr_20,Single_23,Single_23}, i.e., currents flowing through the substrate depend on the magnetization.

Canonical explanation of the CISS effects involves spin-orbit coupling (SOC), and the resulting  Edelstein effect has been investigated in chiral crystals~\cite{PhysRevB.110.045434,chiral_per_zutic_25}. SOC can induce spin-polarized bands in one-dimensional, non-interacting, and time-reversal-invariant multi-band models~\cite{PhysRevB.102.035445}. However, SOC is almost certainly not enough by itself to explain CISS~\cite{Xie2011} since organic compounds with mostly light elements can not be expected to have SOC strong enough to decisively affect transport at room temperature. Consequently, there is a lively theoretical debate about the origin of the CISS effect~\cite{evers_22}. Part of the effort has identified mechanisms that can differentiate between the two SOC bands, so that one of them dominates transport. Dephasing, i.e., flipping spins from the slower to the faster variety, can achieve moderate spin polarization~\cite{PhysRevB.111.205417}, as can the combined impact of SOC and friction~\cite{PhysRevB.104.024430}. Another route focuses on orbital polarization that can be sizable in one-dimensional chiral systems~\cite{chiral_orb_dna,orb_edel_25}, and could be transferred to the spin via SOC of the substrate~\cite{chiral_orb_dna, YANG2025100015}. The substrate itself has been proposed to play a decisive role~\cite{chiral_orb_dna,spin_charge_25}; however, a CISS effect has also been reported for isolated molecules~\cite{CISS_no_substr_23}. Electron-phonon coupling to chiral phonon modes~\cite{doi:10.1021/acs.jpclett.2c03224,doi:10.1126/sciadv.adv5220}, phonon-driven correlations~\cite{PhysRevB.102.235416}, and dissipation~\cite{fransson_T_phonons} have also been discussed.

Finally, electron-electron correlations have been shown to support spin-polarized transport~\cite{fransson_correl_19,Chiesa_24,1d_hubb_23}, but the underlying mechanism has not been clarified. It has recently been pointed out that interaction with non-collinear spin patterns has the symmetries required for CISS~\cite{10.1063/5.0214919,PT_Theiler_26}. We connect these last aspects and investigate non-collinear spin-spin correlations in interacting one-dimensional models, which support spin-polarized transport via the non-relativistic Edelstein effect~\cite{nonrel_edel_25,noncol_alter_25,PhysRevLett.119.187204,2025arXiv250812362P}. In the latter or '$p$-wave magnetism'~\cite{Hellenes2023_pwave}, the coupling of momentum and spin is not due to SOC, but driven by magnetic exchange interactions, which have a much higher energy scale than SOC in light elements.

Using a variety of numerical tools, i.e., density-matrix renormalization group (DMRG), cluster perturbation theory (CPT), and Markov chain Monte Carlo (MCMC) methods, we will show that non-collinear spin spirals arise in several strongly correlated models describing one-dimensional molecules. Moreover, short-range spin-spin correlations will be found to be enough to support signatures of CISS as long as some, even vanishingly small, SOC is present. This rationalizes the experimental observation of CISS at temperatures well above the SOC energy scale. 

\textit{Strongly correlated model ---}
Figure~\ref{fig:cartoon} illustrates a toy model for chiral molecules, similar to Ref.~\cite{chiral_orb_dna}. For simplicity and to permit numerical studies on larger chains, we consider two $p$ orbitals per site, tangential $p_{t}$ and radial $p_{r}$. The Slater-Koster approach~\cite{PhysRev.94.1498} can be used to estimate hopping integrals~\cite{SK_DNA_16}, see End Matter, Appendix~A. In our investigation, we go beyond independent-electron theory and complement this model with on-site interactions comprising Coulomb and Hund's-rule terms, setting the ratio of Hund's-rule coupling $J_{H}$ to Coulomb repulsion $U$ to $J_{H}/U = 1/4$. If the two orbitals have different hopping $|t_{t}/t_{r}|\gtrsim 0.3$ or if they have different onsite energies, they can undergo an orbital-selective Mott phase (OSMP) transition \cite{Georges2013}. In this mixture of metallic and localized bands, the less itinerant $p_{r}$ orbital becomes half-filled, and the more itinerant $p_{t}$ orbital has partial (non-integer) filling. Since this largely suppresses hoppings involving $p_r$, one can further reduce the Hilbert space by considering the corresponding Kondo-lattice model (KLM)~\cite{Herbrych2019,Herbrych2020,Sroda2021,Sroda2023}
\begin{eqnarray}
  H_{\textrm{K}} &=& t_{t} \sum_{i,\sigma} 
  \left(\cdag_{i,t,\sigma}\cnod_{i+1,t,\sigma} + \textrm{H.c.}\right) 
  + U\sum_{i} n_{i,t,\uparrow}n_{i,t,\downarrow}\nonumber\\
  &-& 2J_H \sum_{i} \vec{S}_{i,r}\vec{S}_{i,t} 
  + J_r \sum_{i} \vec{S}_{i,r}\vec{S}_{i+1,r}\,.
  \label{eq:h_blockspiral}
\end{eqnarray}
Here, $\cdag_{i,t,\sigma}$ creates an electron in the tangential orbital $p_{t}$ at site $i$ and with spin $\sigma$, $n_{i,t,\sigma}$ and $\vec{S}_{i,\alpha=t,r}$ are the corresponding density and spin operators. $J_r = 4t_{r}^2/U\ll t_{t}$ effectively takes into account the leading contributions from the suppressed hopping $t_{r}$ of the radial orbital.

The competing double- and superexchange mechanisms within the low-dimensional KLM support a variety of magnetic (quasi-long-range) orders: from "standard" FM and antiferromagnetic through block and block-spiral orders. The latter can be stabilized for Coulomb interactions $U \approx 4 J_{H} \approx 2W$~\cite{BSpiral_20,Han_2025}, where $W$ is the bare kinetic energy bandwidth. Such a non-collinear magnetic order arises from frustration between antiferromagnetic correlations induced by onsite repulsion $U$ and FM double-exchange arising from Hund's coupling $J_H$. While the corresponding one-particle spectrum~\cite{BSpiral_20} somewhat resembles non-interacting bands with large Rashba-SOC, it is important to note that SOC was not included in~\cite{BSpiral_20}, and the band splitting instead came from electronic correlations.

\begin{figure}
  \includegraphics[width=1.0\columnwidth]{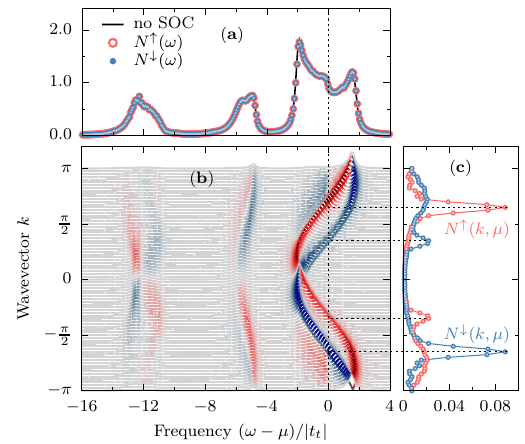}
  \caption{Density matrix renormalization group method results for the KLM~\eqref{eq:h_blockspiral} with $l=1$ Rashba SOC~\eqref{eq:SOC}. (a)~Frequency dependent density-of-states $N^\sigma(\omega)$ without and with Rashba SOC~\eqref{eq:SOC}. For the latter, both spin projections are presented. (b)~Momentum $k$ resolved one-particle spectral density $N(k,\omega)$ for both spins, red and blue represent $\uparrow$ and $\downarrow$ projections of $\vec{d}\cdot\vec{\sigma}$, respectively. (c)~Spectral weight $N(k,\omega=\mu)$ at Fermi level $\mu$. Solid lines in (b) represent the band splitting by the effective (interaction-induced) SOC coupling \mbox{$\omega_\mathrm{eff}(k)=2t_\mathrm{eff}\cos(k)\pm2\lambda_\mathrm{eff}\sin(k)$}, with $t_\mathrm{eff}\simeq0.4$ and $\lambda_\mathrm{eff}\simeq0.25$. Note that $\lambda_\mathrm{eff}$ is an order of magnitude larger than the one used in a full calculation. Parameters used: $t_{t}=-0.5$, $\lambda=0.04\,t_{t}$, $U=8.4\,t_{t}$, $J_{H}=U/4$, $J_r = 0.0428\,t_{t}$ and a $n=3/2$ electron filling of the tangential orbital.}
  \label{fig:Akw_block}
\end{figure}

Chirality also allows for time-reversal invariant SOC of the form 
\begin{align}\label{eq:SOC}
  H_{\textrm{SOC}(l)} = \lambda\sum_j
  (\cdag_{j,t,\uparrow}, \cdag_{j,t,\downarrow} )
  \left( i\vec{d}\cdot\vec{\sigma} \right)
  \left(\begin{array}{c}
  \cnod_{j+l,t,\uparrow}\\ \cnod_{j+l,t,\downarrow}
  \end{array}\right) + \textrm{H.c.}\;,
\end{align}
where $\vec{\sigma}$ is the vector of Pauli matrices and $l$ the hopping distance. In a real-space coil along $z$, $\vec{d}\approx \vec{e}_{z}$ and thus $i\vec{d}\cdot\vec{\sigma} = i \sigma^z$. We use the DMRG method~\cite{Jeckelmann2002,Nocera2016} to obtain the one-particle spectral functions and corresponding density of states for $H= H_{\textrm{K}} + H_{\textrm{SOC}(1)}$, i.e., for nearest-neighbor (NN) SOC $l=1$ and small $\lambda/t_{t}=0.04$. The density of states $N^\sigma(\omega)$ presented in Fig.~\ref{fig:Akw_block}(a) is shown to coincide with that for $\lambda=0$, and comparison with~\cite{BSpiral_20} further confirms that total spin-integrated spectra are not affected by such a small SOC. Figure~\ref{fig:Akw_block}(b) analyzes the $\sigma=\uparrow,\downarrow$-resolved spectra $N(k,\omega)$, i.e., projected onto eigenstates of the Rashba SOC $i\vec{d}\cdot\vec{\sigma}$ in~\eqref{eq:SOC}. One clearly sees strong band splitting and spin polarization of states with $k>0$ (or $k<0$).

Interestingly, the band splitting itself can be reproduced by a large effective $\lambda_\mathrm{eff}/t_{t}\simeq 0.2 \gg 0.04$, see Fig.~\ref{fig:Akw_block}(b). Its magnitude is not controlled by $\lambda$, but by the \mbox{$(U,J_\mathrm{H})$-interaction} and electronic filling \cite{BSpiral_20}. Even more importantly, Rashba SOC acting on a non-interacting band would yield spin-split bands with equal weights, so that one state of each spin projection would move with $k>0$ ($k<0$), and their contributions to transport would largely cancel. Here, in contrast, the 'up' ('down') band is strongly suppressed for $k>0$ ($k<0$), see Fig.~\ref{fig:Akw_block}(c), so that 'down' ('up') dominates transport in positive (negative) direction. See Fig.~\ref{fig:Akw_block}(c) for states at the Fermi level. This would be expected to yield spin polarization of photoelectrons~\cite{Gohler2011,Xie2011} moving through the chiral chain. 

We want to emphasize that already the one-particle spectrum of~\eqref{eq:h_blockspiral} (without SOC) is the sum of the two spin projections, but with an unknown quantization axis~\cite{BSpiral_20}. The role of finite (presumably even infinitesimal) SOC is merely to "choose" the latter. Furthermore, it has recently been pointed out that symmetry breaking into a non-collinear spin pattern can support spin-polarized transport, where the polarized spin component is orthogonal to the spontaneously selected spin plane~\cite{nonrel_edel_25,noncol_alter_25,PhysRevLett.119.187204,2025arXiv250812362P}. 

\textit{Short-range correlations ---}
For organic molecules, long-range magnetic order (or even quasi-long-range) is not expected. We will now show that the importance of SOC (and molecular chirality) lies in ensuring that spin polarization can still be observed even in the presence of merely short-range magnetic correlations. To this end, we will use the CPT method~\cite{PhysRevB.66.075129,Pot03a}, which replaces the self-energy of the entire chain with that of a short segment. Consequently, only short-range electronic correlations are included, see End Matter, Appendix~B.

\begin{figure}
  \includegraphics[width=1.0\columnwidth]{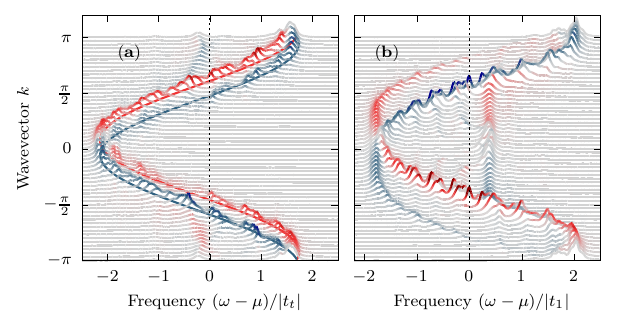}
  \caption{Cluster perturbation theory results based on short-range correlations. Color indicates spin polarization. Panel~(a) is based on an eight-site KLM~\eqref{eq:h_blockspiral} and~\eqref{eq:SOC} with $\lambda =0.015\,t_{t}$, $J_{t}=0.48\,t_{t}$ and $l=1$, remaining parameters as in Fig.~\ref{fig:Akw_block}. Similar as in Fig.~\ref{fig:Akw_block}(b), the solid lines in represent the effective band splitting \mbox{$\omega_\mathrm{eff}(k)=2t_\mathrm{eff}\cos(k)\pm2\lambda_\mathrm{eff}\sin(k)$}, with $t_\mathrm{eff}\simeq0.45$ and $\lambda_\mathrm{eff}\simeq0.1$. Panel~(b) depicts results for a twelve-site Hubbard model combining (\ref{eq:hubbard}) and SOC~\eqref{eq:SOC} with $\lambda=0.01 |t_{1}|$ and $l=2$. Parameters: $t_{1}=-1$, $U=8\,|t_{1}|$, $t_{2}=0.3 \,|t_{1}|$, and $n=2/3$ filling. 
  \label{fig:Akw_8s_cpt}}
\end{figure}

In Fig.~\ref{fig:Akw_8s_cpt}(a), we firstly consider the KLM (\ref{eq:h_blockspiral}) with NN SOC (\ref{eq:SOC}) as before, focusing on the frequencies close to the Fermi level $\mu$. Compared to the spectrum presented in Fig.~\ref{fig:Akw_block}(b), the 'rugged' spectrum reflects the finite length of the correlated cluster, but right- and left-moving bands are still clearly spin polarized. Consequently, $\lambda$ of a few percent of the hopping integrals, i.e., in the realistic meV range, allows for strongly spin-polarized bands even for a system with a short spin-spin correlation length.

\textit{Beyond NN couplings ---}
We have thus seen that chiral spin correlations produce spin-polarized bands when electron-electron interactions are included in a plausible model for chiral molecules. Unfortunately, block-spirals only arise in rather narrow parameter windows of the Kondo-lattice and two-band Hubbard models~\cite{BSpiral_20}, and simple spirals tend not to be stable in KLMs with purely NN terms~\cite{Han_2025}. On the other hand, spirals have been found in models where interactions or hopping act beyond NN~\cite{PhysRevB.79.094425,PhysRevB.58.R11833}, and in various doped multi-orbital systems~\cite{Dong2008,Azhar2017,doped_spin1_26} for which additional (longer range) terms are potentially relevant. The latter couplings were reported to favor the CISS effect~\cite{doi:10.1073/pnas.1407716111} and can be expected for the extended orbitals of organic molecules.

We thus study - as a second example of a correlated model for CISS - a single-orbital Hubbard Hamiltonian
\begin{align}\label{eq:hubbard}
  H_{\textrm{H}} &= \sum_{l=1}^2 t_{l} \sum_{i,\sigma}
  \left(\cdag_{i,\sigma}\cnod_{i+l,\sigma} + \mathrm{H.c.}\right)
  + U\sum_{i} n_{i,\uparrow}n_{i,\downarrow}
\end{align}
with the addition of weak next-NN (NNN) SOC, i.e., (\ref{eq:SOC}) with $l=2$~\cite{Chiesa_24,fransson_correl_19}. We base CPT calculations on a twelve-site chain with eight electrons and $n=2/3$ filling. Compared to the KLM~\eqref{eq:h_blockspiral}, the itinerant electrons here are not coupled to any fully localized spins, but the spectral density in Fig.~\ref{fig:Akw_8s_cpt}(b) nevertheless shows spin-polarized bands.

\textit{Finite temperatures ---} 
A characteristic feature of the CISS effect is that it can be observed at elevated temperatures, above the plausible energy scales for SOC~\cite{Gohler2011,Xie2011,nanofiber_20,fransson_T_phonons}. To obtain easier access to higher temperatures, we consider the KLM with a \textit{classical} localized spin rather than the quantum $\vec{S}_{i,r}$. The classical spin can either model localized spins stemming from strongly correlated half-filled orbitals~\cite{PhysRevB.58.6414,PhysRevB.98.035124,PhysRevLett.88.187001} or can be interpreted as an effective local exchange field (e.g., in a mean-field decoupling~\cite{PhysRevLett.101.156402}). MCMC can then be used without any sign problems~\cite{PhysRevLett.80.845}. In the following, we will consider $H=H_{\textrm{K}}+H_{\textrm{SOC}(2)}$, i.e.~\eqref{eq:h_blockspiral} and~\eqref{eq:SOC} with $l=2$,  with classical localized spins, supplemented with NNN hopping $t_{t,2}$ and NNN Heisenberg coupling $J_{r,2}$ between localized spins, see End Matter, Appendix~C.

\begin{figure}
  \includegraphics[width=1.0\columnwidth]{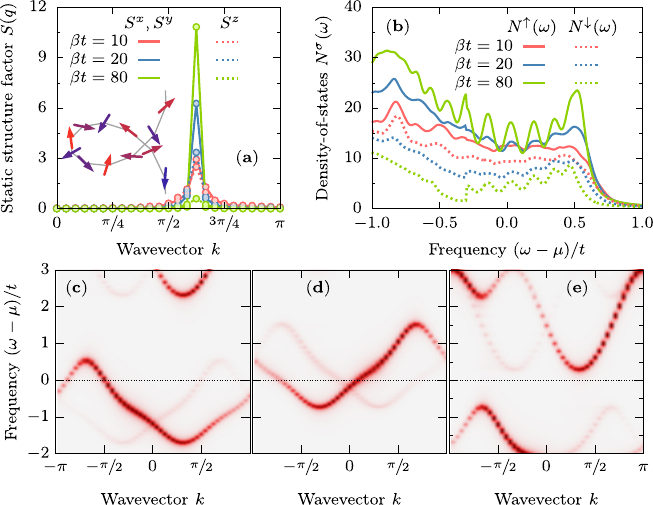}
  \caption{Markov-chain Monte-Carlo results for the classical spin KLM. (a) Spin-structure factor $S(k)$ for in-plane ($x$, $y$) and out-of-plane ($z$) components of spin and various temperatures. Parameters: NN and NNN hoping $t_{t}=-1$, $t_{t,2}=0.25\,|t_{t}|$, respectively, NN and NNN Heisenberg spin exchange $J_r=0.8\,|t_{t}|$, $J_{r,2} = 0.5\,|t_{t}|$, respectively, coupling to localized spins $2J_H|S_{i,r}|=2\,|t_{t}|$, $n=0.8$ filling, and SOC $\lambda = 0.02\,|t_{t}|$, $l=2$. (b) Density of states $N^\sigma(\omega)$ for left-moving states for parameters as in (a). (c) Momentum $k$ resolved one-particle spectral density $A(\vec{k},\omega)$ for $\sigma=\uparrow$ at $\beta\,|t_{t}|=80$, remaining parameters as in (a). (d) $A(\vec{k},\omega)$ for $\sigma=\uparrow$ for filling $n=0.4$, remaining parameters as in (c). (e) $t_{t,2} = 0.25\,|t_{t}|$, $J_r = 0.8\,|t_{t}|$, $J_{r,2} = 0.5\,|t_{t}|$, $2J_H|S_{i,r}|=1.5\,|t_{t}|$, $n=1$ and $\lambda = 0.02\,|t_{t}|$, $l=2$.} 
  \label{fig:MCMC_SOC2}
\end{figure}

The NNN terms frustrate the lattice and induce a spiral. The inset of Fig.~\ref{fig:MCMC_SOC2}(a) shows part of an MCMC snapshot, and the static spin-structure factor $S(k)$ is peaked at incommensurate $k\lesssim 2\pi/3$. Next, small SOC $\lambda = 0.02\,|t_t|$ pushes spins in the $x$-$y$ plane and selects the spiral's winding direction. Both long-range order and in-plane preference are lost with rising temperature, but this temperature scale is set by the competition of $J_{r,2}$ and $t_{t,2}$, and can become much larger than $\lambda$. Accordingly, the density of states $N^\sigma(\omega)$ [presented for left-moving states in Fig.~\ref{fig:MCMC_SOC2}(b)] also shows substantial spin polarization well above $T\approx \lambda$. 

\textit{Relation to experiment ---}
States "moving" in a given direction have a preferred spin character, explaining spin-selective propagation of photo electrons~\cite{Gohler2011,Xie2011}. 
All our examples are for right-handed spirals, and yet spins are 
polarized parallel to the electron’s momentum in some cases, and anti-parallel
in others. Since the spin-spin angle determining polarization is not set by SOC directly, but
arises from its interplay with correlations, even changing band filling can switch polarization, see Fig.~\ref{fig:MCMC_SOC2}(c,d). Indeed, opposite spin-chirality relations have been found in peptides w.r.t. DNA~\cite{peptide_dna_mr_20,peptide_mr_16,Xie2011}.  

Most band structures shown above would at first sight suggest that a current can more easily flow out of a substrate with 'down' polarization and into one with opposite 'up' polarization, at least for small voltages. In CISS systems, however, currents in either direction (into and out of the substrate) are higher for a specific magnetization direction, which changes with the molecules' chirality~\cite{nanofiber_20,peptide_dna_mr_20,peptide_mr_16,Xie2011,Single_23,Fransson_25}. The debated~\cite{EMChA_vs_CISS_23} magnetoresistance is seen over a wide energy range, and a non-equilibrium treatment of our strongly correlated systems is beyond the scope of this work.  

Yet we want to point out two mechanisms that are valid near equilibrium, provided spin correlations remain intact. First, gaped (i.e., insulating or semi-conducting) bands as in Fig.~\ref{fig:MCMC_SOC2}(e) support the transport asymmetry seen in CISS: highest occupied states with $k< 0$ have dominant 'up' polarization, as do the lowest empty states with $k>0$. Charge flow in either direction thus favors the same polarization. 

Second, we would like to highlight the role of the substrate in the case of metallic molecules. When chiral molecules approach a surface, some charge is likely to transfer into or out of the molecule, and either direction of charge flow has a preferred spin polarization. A molecule of a given chirality can then be expected to attach more easily if the substrate states involved in charge transfer have the matching spin, which leads to the observed enantiomer-specific adsorption~\cite{Adsorb_Ernst_24,deposition_18}~\footnote{The associated transient current would additionally induce a magnetic moment due to the (here largely non-relativistic) Edelstein effect, but the interaction of this moment with the substrate magnetization is likely a secondary effect~\cite{chiral_orb_dna}.} The quality of the binding would, in turn, affect charge transport. 

\MD{Moreover}, in FM substrates like Ni, Co, or Fe, specific subbands of $t_{2g}$ or $e_g$ character have a much more pronounced spin polarization than their average would suggest~\cite{PhysRevB.76.035107,FeNi_DMFT_17}. A good match of $d$-subband energy with molecular states has been found to have an important impact on adsorption, even taking it from physisorption to chemisorption~\cite{bonds_ads_25}. Such a stronger role of specific $d$ states might thus explain why CISS magneto-resistance can become stronger~\cite{Fransson_25} than the overall, i.e., averaged, spin polarization of the substrate.

\textit{Conclusions ---}
We have shown here that models that qualitatively describe organic molecules support noncollinear spin-spin correlations when electron-electron interactions are included. While a coplanar spiral order does not break time-reversal invariance, it does break parity. Electronic bands in such $p$-wave magnets then obey 
\begin{align}\label{eq:pwave_tp}
  \epsilon_\uparrow(\vec{k}) =
  \epsilon_\downarrow(-\vec{k}),\quad\textrm{but}\quad
  \epsilon_\sigma(\vec{k}) \neq \epsilon_\sigma(-\vec{k})\;,
\end{align}
where spin $\sigma$ refers to a quantization axis perpendicular to the spin-plane of the spiral~\cite{nonrel_edel_25,noncol_alter_25,PhysRevLett.119.187204,2025arXiv250812362P}. As shown above, bands with positive/negative momentum thus become spin-polarized. 

The role of SOC, tied to the molecules' chirality, is to ensure that short-range-ordered segments in disordered and imperfect systems have aligned chiralities. The energy cost of flipping a spiral segment with $N$ correlated sites into the opposite chirality scales with $N \lambda$, thereby allowing spin polarization to persist at temperatures far exceeding the energy scale of bare SOC and rationalizing the observation of correlation-enhanced CISS~\cite{Chiesa_24}. 

Future theoretical work on CISS should thus take into account electron-electron correlations beyond simple exchange splittings. In particular, the interplay between noncollinear spin-spin correlations and chiral vibrations or orbital polarization is expected to further clarify the origin of CISS.

\textit{Acknowledgments}
J.H. acknowledges grant support by the National Science Centre (NCN), Poland, via Sonata BIS project no. 2023/50/E/ST3/00033. Part of the calculations have been carried out using resources provided by the Wroclaw Center for Networking and Supercomputing (\url{http://wcss.pl}). M.D. gratefully acknowledges helpful discussions with S. Loth and H. Zacharias. The data that support the findings of this article are openly available \cite{opendata}.

\bibliography{spirals}

\clearpage

\onecolumngrid
\section*{End Matter}
\twocolumngrid

\renewcommand{\theequation}{A\arabic{equation}}
\setcounter{equation}{0}
\textbf{ Appendix A: Slater-Koster Approach ---}
The Slater-Koster (SK) approach is based on hopping amplitudes tabulated for the $p_x$ and $p_y$ orbitals~\cite{PhysRev.94.1498}. Their positions are given by cylindrical coordinates $(x_{i},y_{i},z_{i}) = (\rho\cos \phi_{i}, \rho \sin \phi_{i},z_{i})$. The spiral is defined by the increments of $\phi_{i}=i\phi=i2\pi/n$ and $z_{i}=i\Delta z$ between sites. This yields hoppings between $p_x$ and $p_y$ orbitals at different sites. We then translate them into the basis of  tangential and radial orbitals via  
\begin{eqnarray}
 \begin{pmatrix}
 |p_{r,i}\rangle\\
 |p_{t,i}\rangle
 \end{pmatrix}=
 \begin{pmatrix}
 \cos \phi_{i} & \sin \phi_{i}\\
 -\sin \phi_{i} & \cos \phi_{i}
 \end{pmatrix}
 \begin{pmatrix}
 |p_{x,i}\rangle\\
 |p_{y,i}\rangle
 \end{pmatrix}\,
\end{eqnarray}
to recover translational invariance. Using these basis transformations and the directional cosines following from lattice coordinates, hoppings can then be expressed in terms of Slater-Koster parameters $(pp\sigma)$ and $(pp\pi)$.

Typically, $(pp\sigma) < 0 $ and $|(pp\sigma)| > (pp\pi) > 0$. For DNA, $\Delta z/\rho\approx 1/3$ and $n\approx 10$. We then find that ratios $(pp\pi)\lesssim 0.15 |(pp\sigma)|$ give small hopping of the radial orbital, i.e., $|t_{r}/t_{t}|<1/3$. For $(pp\pi)\gtrsim |(pp\sigma)|/2$, on the other hand, the role of $p_t$ and $p_r$ switch with $|t_{r}/t_{t}|> 3$, leading to analogous results. In either case, $t_{r,t}=-t_{t,r}$ are small, so that the conditions for an OSMP are fulfilled for orbitals with degenerate onsite energies. A difference in onsite energies can significantly extend the range of the OSMP, as band fillings then become unequal already without hopping. It is worth noting that the OSMP is not strictly necessary to find spin spirals, see Fig.~\ref{fig:Akw_8s_cpt}(b) of the main text.

The on-site interactions for the case of two orbitals are given by the so-called Hubbard-Kanamori model
\begin{align}
  H_{\textrm{int}} &= U \sum_{i, \alpha} n_{i \alpha \uparrow} n_{i \alpha \downarrow}  \label{Hint}
  +\frac{U^\prime}{2} \sum_{i, \sigma} \sum_{\alpha \neq \beta} n_{i
  \alpha \sigma} n_{i \beta \bar{\sigma}}\\ \nonumber
  & +\frac 1 2 (U^\prime - J_H) \sum_{i,\sigma} \sum_{\alpha \neq
  \beta} n_{i \alpha \sigma} n_{i \beta \sigma}\\ \nonumber
  & + J_H \sum_{i, \alpha \neq \beta} (\cdag_{i \alpha \uparrow}
  \cdag_{i \alpha \downarrow} \cnod_{i \beta \downarrow} \cnod_{i \beta
  \uparrow}
  - \cdag_{i \alpha \uparrow} \cnod_{i \alpha \downarrow} \cdag_{i, \beta \downarrow} \cnod_{i \beta \uparrow})
\end{align}
where $\alpha=t,r$ indicates the orbital, and $i$ the site. Coulomb interaction $U$, $U^\prime$ and Hund's coupling $J_{H}$ are connected via $U^\prime = U-2J_{H}$~\cite{PhysRevB.18.4945,PhysRevB.28.327}. Plausible orders of magnitude for the largest NN hoppings were found to be in the range of $\approx 100\,\textrm{meV}$~\cite{doi:10.1021/jp401705x}. At $U=8t_{t}$ and $J_H/U=1/4$, this would give a local singlet-triplet splitting $2J_{H} \approx 4 t_{t} \approx 0.4\;\textrm{eV}$ of the order of half an eV. This is somewhat smaller than but comparable to estimates for organic polymer chains~\cite{QuChem_ST_poly_04}.

\renewcommand{\theequation}{B\arabic{equation}}
\setcounter{equation}{0}
\textbf{Appendix B: Cluster Perturbation Theory ---}
The one-particle spectral density is given by the imaginary part of the one-electron Green's function, i.e.,
\begin{align}\label{eq:Nkw}
  N^\sigma(k,\omega) &=  -\frac{1}{\pi L}\textrm{Im}\langle \textrm{GS} |\cdag_{k,\sigma}
  \frac{1}{\omega - (H-E_0) + i\eta}
  \cnod_{k,\sigma}|\textrm{GS}\rangle\nonumber\\
  &\quad -\frac{1}{\pi L}\textrm{Im}\langle \textrm{GS} |\cnod_{k,\sigma}
  \frac{1}{\omega + (H-E_0) + i\eta}
  \cdag_{k,\sigma}|\textrm{GS}\rangle
\end{align}
obtained for the ground state $|\textrm{GS}\rangle$ of an $L$-site chain. The density of states corresponds to the sum over momenta
\begin{align}\label{eq:dos}
  N^\sigma(\omega) &= \frac{1}{L} \sum_{k} N^\sigma(k,\omega)\;.
\end{align}

In CPT, the one-particle Green's function (\ref{eq:Nkw}) is obtained in a two-step process. Firstly, the cluster Green's function $G_{\textrm{Cl}}$ with matrix elements 
\begin{align}\label{eq:Gcl}
  G_{\textrm{Cl},ab}(\omega) &=&  \langle \textrm{GS} |\cdag_{a}
  \frac{1}{\omega - (H-E_0) + i\eta}
  \cnod_{b}|\textrm{GS}\rangle\\
  & +& \langle \textrm{GS} |\cnod_{a}
  \frac{1}{\omega + (H-E_0) + i\eta}
  \cdag_{b}|\textrm{GS}\rangle \nonumber
\end{align}
is calculated for a small cluster with open boundaries using exact diagonalization (here with the Lanczos algorithm). Indices $a,b$ combine site, orbital, and spin. Secondly, inter-cluster hopping $T_{\textrm{Cl-Cl}}$ between the small clusters is added as a perturbation~\cite{PhysRevB.66.075129}
\begin{align}\label{eq:CPT_1}
  G_{\textrm{CPT}}^{-1} = G_{\textrm{Cl}}^{-1} + T_{\textrm{Cl-Cl}} 
\end{align}
when obtaining the CPT Green's function
$G_{\textrm{CPT}}$. Introducing $G_{\textrm{0,Cl}}^{-1}$ for the
inverse non-interacting Green's functions of the cluster and
$G_{0}^{-1} = G_{\textrm{0,Cl}}^{-1} + T_{\textrm{Cl-Cl}}$ for that of
the full system, one can then find that  
\begin{align}\label{eq:CPT_2}
  G_{\textrm{CPT}}^{-1} = G_{\textrm{Cl}}^{-1} -
  G_{\textrm{0,Cl}}^{-1} + G_{\textrm{0}}^{-1} =
  \Sigma_{\textrm{Cl}} + G_{\textrm{0}}^{-1}\;.
\end{align}
Here, the self-energy $\Sigma_{\textrm{Cl}}$ is the difference between non-interacting and interacting Green's functions of the small cluster, and thus captures the impact of correlations. CPT replaces the self-energy of the full system by that of a small cluster~\cite{Pot03a}, thereby reducing the impact of electron-electron correlations to the short distances of the fully solved segment.  

In the ED step of the CPT calculations for Fig.~\ref{fig:Akw_8s_cpt}(a), states with doubly-occupied orbitals $p_t$ were excluded to reduce the numerical effort. To partially make up for neglected superexchange effects, we included an explicit NN spin-spin coupling $J_t \vec{S}_{i,t}\vec{S}_{i+1,t}$ between itinerant electrons, but only acting within each small eight-site cluster.

\renewcommand{\theequation}{C\arabic{equation}}
\setcounter{equation}{0}
\textbf{Appendix C: Markov-chain Monte-Carlo method ---}
In the classical MCMC algorithm, the key approximations to (\ref{eq:h_blockspiral}) are (i) that the spin degrees of freedom are considered classical and (ii) that the Hubbard interaction between itinerant electrons is neglected. Since the itinerant spins are coupled ferromagnetically to the localized spin, the second approximation becomes less drastic: electrons with the wrong spin polarization are already pushed to higher energy, so that (at most) single occupancy is energetically favored~\cite{PhysRevB.58.6414,PhysRevB.98.035124,PhysRevLett.88.187001}.

Each spin configuration is thus represented by a chain of $L$ vectors $\vec{S}_{i,r}$ of unit length, i.e., each parameterized by two angles $\theta_i$ and $\phi_i$. These spins then define an effective non-interacting electronic Hamiltonian
\begin{eqnarray}
  H_{\textrm{eff}}(\{\theta_i,\phi_i\})
  &=& t_{t} \sum_{i,\sigma}
  \left(\cdag_{i,t,\sigma}\cnod_{i+1,t,\sigma} + \textrm{H.c.}\right)  \\
  &+& t_{t,2} \sum_{i,\sigma}
  \left(\cdag_{i,t,\sigma}\cnod_{i+2,t,\sigma} + \textrm{H.c.}\right) \nonumber\\ 
  &-& 2J_H \sum_{i} \vec{S}_{i,r}\vec{S}_{i,t} \nonumber\\
  &+& J_r \sum_{i} \vec{S}_{i,r}\vec{S}_{i+1,r} + J_{r,2} \sum_{i} \vec{S}_{i,r}\vec{S}_{i+2,r}\nonumber\,. 
\end{eqnarray}
This effective one-particle Hamiltonian can be diagonalized efficiently, and thermodynamic expectation values can be easily evaluated (the free energy of the electrons enters the acceptance probability of the MCMC algorithm). Temperatures were lowered in gradual steps to facilitate reaching the equilibrium state. At each temperature, step sizes were adjusted to reach acceptance rates of $\approx 40\%$ during equilibration. Runs were repeated with different seeds, and care was taken to ensure that the results no longer depended on simulation parameters such as the number of MCMC steps skipped between measurements.

Observables are calculated from MCMC spin configurations, e.g. the $\alpha =x,y,z$ component $S^\alpha(k)$ of the spin-structure factor 
\begin{align}
  S^\alpha(k) = \frac{1}{L} \Bigl| \sum_j \e^{ikj} S^\alpha_{j,r} \Bigr|^2\;,
\end{align}
and then averaged over configurations. The spectral density (\ref{eq:Nkw}) is obtained from the eigenvectors of the effective electronic Hamiltonian for each spin configuration entering the average. For the left-moving density-of states shown in Fig.~\ref{fig:MCMC_SOC2}(b), the sum (\ref{eq:dos}) is only taken over states with $\partial_k \epsilon_k < 0$. Such a case is found for momenta $-3\pi/4\lesssim k_z\lesssim 0$ (mostly spin $\uparrow$) resp. $3\pi/4\lesssim k_z \pi$ (mostly spin $\downarrow$). The wiggles observed at the lowest temperature are a finite-size effect arising from limited momentum resolution. 

\end{document}